\def\lang#1#2{#1}

\documentclass[12pt]{article}
\usepackage[utf8]{inputenc}
\usepackage[T1]{fontenc}
\usepackage{cmap,longtable}
\lang{}{\usepackage[ukrainian,english]{babel}}

 \def\b{\beta}
  \def\k{\kappa}\def\L{\Lambda}
 \def\s{\sigma} \def\th{\theta}\def\om{\omega}

 \def\P{\Psi} \def\ph{\phi}

\def\imo{i}

\def\be{\begin{equation}}
\def\ee{\end{equation}}
\def\bea{\begin{eqnarray}}
\def\eea{\end{eqnarray}}

\newcommand{\p}[1]{(\ref{#1})}

\begin{document}
\lang{}{\selectlanguage{ukrainian}\cyr}
\thispagestyle{empty}
\centerline{\bf \Large \lang{Quasi-normal modes of Schwarzschild-de Sitter black holes}{Квазинормальні моди чорних дір Шварцшильда-де Сітера}}

\bigskip

\centerline{\bf \lang{A. Zhidenko}{О. Жиденко}}

\centerline{\lang{Department of Physics, Dniepropetrovsk National University}{Дніпропетровський Національний Університет, фізичний факультет}}
\centerline{\lang{St. Naukova 13, Dniepropetrovsk 49050, Ukraine}{вул. Наукова, 13, Дніпропетровськ 49050, Україна}}
\vskip3mm
\bigskip \nopagebreak

\begin{abstract}
\lang{The low-laying frequencies of characteristic quasi-normal modes (QNM) of Schwarzschild-de Sitter (SdS) black holes have been calculated for fields of different spin using the sixth-order WKB approximation and the approximation by the Pöshl-Teller potential. The well-known asymptotic formula for large $l$ is generalized here on a case of the Schwarzschild-de Sitter black hole. In the limit of the near extreme $\L$ term the results given by both methods are in a very good agreement, and in this limit fields of different spin decay with the same rate.}
{Обчислено нижні частоти характеристичних квазинормальних мод (quasi-normal modes) чорних дір Шварцшильда-де Сітера (Schwarzschild-de Sitter) для полів різного спіну, використовуючи наближення шостого порядку ВКБ (WKB) і наближення потенціалом Пьошля-Теллера. Узагальнено відому асимтотичну формулу для великих $l$ на випадок чорної діри Шварцшильда-де Сітера. Коли $\L$-член прямує до екстремального значення, результати обох методів, добре узгоджуються між собою, а поля різних спінів згасають з однаковою швидкістю.}
\end{abstract}

\newpage\setcounter{page}1
\section{\lang{Introduction}{Вступ}}
\lang{Within general relativity it is well-known that there are three stages to the dynamical evolution of field perturbation on a black hole background: the initial outburst from the source of perturbation, the damping (quasi-normal) oscillations and asymptotic tails at very late time. The evolution significantly depends on the asymptotic behavior of the background: for different values of cosmological constant $\L$ there are three types of behavior depending on the sign of $\L$. Thus, the remarkable fact is that while for asymptotically flat backgrounds we have inverse power law tails, for asymptotically Anti-de Sitter background it is exponential low that governs the decay of a field at all times. The same exponential decay at very late times was observed by Brady \textit{et al} \cite{Brady-Chambers} for asymptotically de Sitter background.}
{З загальної теорії відносності добре відомо, що існує три стадії динамічної еволюції збурення поля на тлі чорної діри: початкова дія джерела збурення, (квазинормальні) коливання, що спадають, та асимптотичні хвости на пізніх етапах. Еволюція істотно залежить від асимптотичної поведінки тла: залежно від знаку космологічної сталої $\L$ існує три типи поведінки. Вартим уваги є факт, що для асимптотично плоского тла асимптотичні хвости є степеневими, а для від'ємної космологічної сталої існує експоненціальне спадання на всіх стадіях. Таке саме експоненціальне спадання на дуже пізніх етапах отримали Брейді та ін. \cite{Brady-Chambers} для додатного $\L$.}

\lang{Apart from original interest coming from possibility to observe quasi-normal ringing with the help of gravitational wave detectors (see, e.g., \cite{QNMrev} for review), the quasi-normal modes bring now a lot of interest in different contexts: in Anti-de Sitter(de Sitter)/Conformal Field Theory correspondence \cite{Horowitz-Hubeny,CFT-Correspond,Moss-Norman}, when considering thermodynamic properties of black holes in loop quantum gravity \cite{LQG}, in the context of possible connection with critical collapse \cite{Horowitz-Hubeny,BHCC}.}
{Крім можливості спостереження квазинормального дзвону з допомогою детекторів гравітаційних хвиль (див. наприклад огляд \cite{QNMrev}), зараз також існує інтерес до квазинормальних мод з інших причин: через існування відповідності між простором анти-де Сітера (де Сітера) та Конформною теорією поля (Conformal Field Theory) \cite{Horowitz-Hubeny,CFT-Correspond,Moss-Norman}, в контексті вивчення термодинамічних властивостей чорних дір у петельній квантовій гравітації, а також через можливе існування зв'язку з критичним колапсом \cite{Horowitz-Hubeny,BHCC}.}

\lang{The gravitational QNMs of Schwarzschild-de Sitter black hole were considered in \cite{Otsuki,Mellor-Moss,Moss-Norman}. Recently, asymptotic values of QNMs were studied for nearly extremal Schwarzschild-de Sitter black hole and it was shown that in the near extremal limit the effective potential of the wave equation reduces to the Pöshl-Teller potential, and it is expected therefore the Pöshl-Teller method provides the best accuracy near the extremal value of $\L$ \cite{Cardoso-Lemos2}.}
{Гравітаційні квазинормальні моди чорної діри у просторі Шварцшильда-де Сітера розглядалися в \cite{Otsuki,Mellor-Moss,Moss-Norman}. Нещодавно при вивченні асимтотичних значень квазинормальних мод біля екстремальної космологічної сталої було показано, що ефективний потенціал хвильового рівняння зводиться до потенціалу Пьошля-Теллера, тому очікується, що метод Пьошля-Теллера дає найкращу точність біля екстремального значення $\L$ \cite{Cardoso-Lemos2}.}

\lang{QNMs corresponding to decay of fields of different spin are extensively studied for Schwarzschild and Schwarzschild-anti de Sitter backgrounds. At the same time there is lack of such study of the Schwarzschild-de Sitter case. This motivated us to calculate QNMs of scalar, electromagnetic, gravitational and Dirac fields for Schwarzschild-de Sitter black hole. We also compare the results found here with those obtained recently in \cite{Cho,Cardoso-Lemos2}.}
{Квазинормальні моди полів різного спіну були докладно вивчені на тлі метрики Шварцшиильда та Шварцшильда-анти-де Сітера, тоді як випадок де Сітера залишається малодослідженим. Ця робота присвячена обчисленню квазинормальних мод скалярного, електромагнітного, гравітаційного та поля Дірака для чорної діри Шварцшильда-де Сітера, а також порівнянню з нещодавно отриманими результатами \cite{Cho,Cardoso-Lemos2}.}

\lang{In section \ref{eq} we give the perturbation equations to describe fields of different spin in SdS background. Section \ref{calc} presents comparison of QNM frequencies obtained through sixth order WKB formula and approximation by the Pöshl-Teller potential.}
{У розділі \ref{eq} наводяться рівняння для опису збурення полів різного спіну на тлі метрики Шварцшильда-де Сітера. Розділ \ref{calc} присвячений порівнянню квазинормальних частот, отриманих з формули ВКБ шостого порядку і наближенням потенціалом Пьошля Теллера.}

\lang{There was shown that the frequencies of the massless Dirac field perturbations \cite{Cho} are the same for both chiralities in SdS background as well. Near the extremal value of $\L$ the sixth order WKB formula gives the values which are very close to those obtained within the Pöshl-Teller potential approximation. This agrees with recent work of Cardoso and Lemos \cite{Cardoso-Lemos2}.}
{Показано, що частоти безмасового поля Дірака \cite{Cho} є також однаковими для обох спіральностей (chirality) у просторі Шварцшильда-де Сітера. Біля екстремального значення $\L$ шостий порядок ВКБ дає значення дуже близькі до тих, які отримані з наближення потенціалом Пьошля-Теллера, що узгоджується з роботою Кардосо й Лемоша \cite{Cardoso-Lemos2}.}

\section{\lang{Perturbation equations and methods}{Рівняння і методи опису збурень}}\label{eq}
\lang{It is well known that the Schwarzschild-de Sitter black hole of mass $M$ is described by the metric}
{Добре відомо, що чорна діра Шварцшильда-де Сітера маси $M$ описується метрикою}:
\be\label{sds}
ds^2 = f(r)dt^2 - \frac{dr^2}{f(r)} - r^2 d\s^2; \qquad f(r) = 1 -
\frac{2M}{r} - \L\frac{r^2}{3}, \quad d\s^2 = d\th^2 + \sin^2\th
d\ph^2.
\ee
\lang{A perturbation equation on this background can be reduced to the Schrödinger wave-like equations}
{Рівняння збурення на такому тлі можна звести до рівняння типу хвильового}:
\be\label{wleq}
\left(\frac{d^2}{dr^{*2}} + \om^2 - V(r^*)\right)\P(r^*) = 0,
\ee
\lang{where we assume exponential dependence of}{де ми припускаємо експоненціальну залежність} $\P$ \lang{on time}{від часу} ($\sim e^{-\imo\om
t}$), $\om = \om_{Re} - i\om_{Im}$, $r^*$ \lang{is the ``tortoise''
coordinate}{``черепашина'' (``tortoise'') координата}:
\be\label{tortoise}
dr^* = \frac{dr}{f(r)}.
\ee
\lang{Under the positive real part QNMs, by definition, satisfy the following boundary conditions}
{Квазинормальні моди з додатною дійсною частиною за означенням задовольняють граничні умови}:
\be\label{bounds}
\P(r^*) \sim C_\pm \exp(\pm\imo\om r^*), \qquad r\longrightarrow
\pm\infty,
\ee
\lang{which correspond to purely out-going waves at infinity and in-going waves at the horizon}
{що відповідає тільки хвилям, які виходять на космологічний горизонт і падають на горизонт подій}.

\lang{The corresponding perturbation equations can be rewritten in form \p{wleq} with potentials}
{Відповідні рівняння можна переписати у формі \p{wleq} з потенціалами}
\bea\label{sp}
V_s &=& f(r)\left(\frac{l(l+1)}{r^2} + \frac{1}{r}f'(r)\right);
\\\label{emp}
V_{el} &=& f(r)\frac{l(l+1)}{r^2};
\\\label{gpo}
V_{odd} &=& f(r)\left(\frac{l(l+1)}{r^2} - \frac{6M}{r^3}\right);
\\\label{gpe}
V_{even} &=& \frac{2f(r)}{r^3}\frac{9M^3 + 3c^2Mr^2 + c^2(1 +
c)r^3 + 3M^2(3cr - \L r^3)}{(3M + cr)^2}, \quad c =
\frac{l(l+1)}{2} - 1;
\\\label{mdp}
V_d &=& \frac{\k f^{1/2}(r)}{r^2}(\k f^{1/2}(r) \pm (\frac{3M}{r}
- 1));
\eea
\lang{for massless scalar, electromagnetic, gravitational (odd end even parities) and Dirac fields respectively (see, e.g., \cite{Potentials} and references therein), where $l$ is the multipole number, $\k = l+1$ for (+) sign and $\k = l$ for (-) sign in \p{mdp}. All the potentials vanish at the event horizon and the cosmological horizon, having a peak near the event horizon. It allows to use the Pöshl-Teller potential approximation as well as WKB method.}
{відповідно для безмасових скалярного, електромагнітного, гравітаційного (аксіального та полярного) і Діраківського полів (див., наприклад, \cite{Potentials} і посилання в ній), де $l$ мультипольне число, $\k = l+1$ для знаку (+) і $\k = l$ для знаку (-) у \p{mdp}. Усі потенціали прямують до нуля на горизонті подій і на космологічному горизонті, маючи максимум біля горизонту подій. Ці властивості дозволяють використовувати і метод наближення потенціалом Пьошля-Теллера, і метод ВКБ.}

\lang{The WKB approach is based on the analogy with the problem of waves scattering near the peak of the potential barrier $V(r^*)$ in quantum mechanics, where $\om^2$ plays a role of energy. The approach was used by Schutz and Will \cite{SchutzWill}, developed to the third order by Iyer and Will \cite{IyerWill}, and recently developed by Konoplya to the sixth
order beyond the eikonal approximation \cite{Konoplya}. The result has the form:}
{Підхід ВКБ виходить з подібності до квантовомеханічної задачі про розсіяння хвиль на потенціальному бар'єрі $V(r^*)$, де роль енергії грає квадрат частоти $\om^2$. Формула ВКБ використовувалася ще Шуцем та Вілом \cite{SchutzWill}, була обчислена Аєром і Вілом до третього порядку \cite{IyerWill}, а нещодавно Р. Коноплею було отримано розвинення поза наближенням ейконала до шостого порядку \cite{Konoplya}. Результат має вигляд:}
\be\label{WKB}
\imo\frac{\om^2 - V_0}{\sqrt{-2V_0^{\prime\prime}}} - \L_2 - \L_3 - \L_4 -
\L_5 - \L_6 = n + \frac{1}{2},
\ee
\lang{where $V_0$ is the height and $V_0^{\prime\prime}$ is the second derivative with respect to the tortoise coordinate of the potential at the maximum. $\L_2$ and $\L_3$ can be found in \cite{IyerWill}, $\L_4$, $\L_5$ and $\L_6$ are presented in \cite{Konoplya}; the corrections depend on the value of the potential and higher derivatives of it at the maximum.}
{де $V_0$ -- значення, а $V_0^{\prime\prime}$ -- друга похідна по черепашиній координаті потенціала в максимумі. $\L_2$ та $\L_3$ можна знайти в \cite{IyerWill}, $\L_4$, $\L_5$ і $\L_6$ виписані в \cite{Konoplya}. Ці поправки залежать від значення потенціала та його вищих похідних у максимумі.}

\lang{Generally, WKB method was effectively used for finding quasi-normal frequencies in a lot of papers \cite{WKB}. Comparing the values calculated by the different order WKB formula one can judge about the convergence to the some unknown accurate result as the WKB order increasing. We expect that the possible error is less then the difference between the sixth and the fifth order WKB values\footnote{The accuracy of values are to take as a half of the number last order (e. g. $0.125$ means $0.125\pm 0.003$), but it is much better for most of them.}. Note, that standard machine precision in MATHEMATICA gives large errors due to numerical truncations. To exclude such ``noises'' we had to hold 40 digits in all intermediate values.}
{Взагалі, метод ВКБ ефективно використовувався для знаходження квазинормальних частот у багатьох роботах \cite{WKB}. Порівнюючи значення, обчислені за допомогою різних порядків ВКБ, можна оцінити наскільки швидко вони збігаються до свого точного значення зі збільшенням порядку метода. Очікується, що можлива похибка є меншою за різницю між шостим і п'ятим порядком ВКБ\footnote{Точність наведених чисел дорівнює половині останнього порядку (наприклад $0.125$ означає $0.125\pm 0.003$), але в більшості випадків вона значно вище.}. Слід зауважити, що стандартна машинна точність програми MATHEMATICA дає велику помилку через відкидання знаків нижніх порядків, тому необхідно утримувати 40 знаків у всіх проміжних значеннях.}

\lang{For comparison we will use the semi-analytical method for calculations proposed by B. Mashhoon \cite{FerrariMashhoon} who used Pöshl-Teller approximate potential}
{Для порівняння використаємо метод обчислення, запропонований Б. Машуном \cite{FerrariMashhoon}, який наближено використовував потенціал Пьошля-Теллера}
\be\label{PT}
V_{PT} = \frac{V_0}{\cosh^2(r^*/b)}.
\ee
\lang{It contains two free parameters ($V_0$ and $b$) which are used to fit the height and the second derivative of the potential $V(r^*)$ at the maximum \cite{FerrariMashhoon}.}
{Цей потенціал має два вільних параметри ($V_0$ і $b$), що визначаються за допомогою висоти і другої похідної потенціалу $V(r^*)$ в максимумі \cite{FerrariMashhoon}.}

\lang{The quasi-normal modes of the Pöshl-Teller potential can be evaluated analytically:}
{Квазинормальні моди для потенціалу Пьошля-Теллера знаходяться аналітично:}
\be\label{PTqnm}
\om = \frac{1}{b}\left(\sqrt{V_0b^2 - \frac 1 4} -(n + \frac{1}{2})\imo\right).
\ee

\section{\lang{QNMs of SdS black holes}{Квазинормальні моди чорних дір Шварцшильда-де Сітера}}\label{calc}
\lang{Both methods give best result for low overtone number but it is known that the higher modes give larger values of the imaginary part of frequencies \cite{QNMrev}. Therefore these modes decay faster then low-lying ones that are more interesting for us. So we calculated some first quasi-normal mode frequencies for massless scalar, electromagnetic and gravitational and Dirac fields using both of the methods. The results are presented below (all values are measured in the black hole mass (M) units).}
{Обидва методи дають кращі результати для нижніх овертонів, проте відомо, що вищі моди мають більші уявні частини \cite{QNMrev}. Тому ці моди спадають швидше і є менш цікавими для даного розгляду. В роботі наведені кілька перших квазинормальних частот для безмасових скалярного, електромагнітного, гравітаційного та Діраківського полів, обчислені обома методами (всі числа подані в одиницях маси (M) чорної діри).}
\def\MSF{\lang{QNMs of massless scalar field}{Частоти скалярного поля}}
\bigskip\newline
\begin{minipage}[c]{9cm}
\centerline{\MSF: $l=1$, $n=0$}
\begin{longtable}{|r|r|r|}
\hline
$\L$ & $\om$ (WKB) & $\om$ (P-T)\\
\hline
   $0$&$0.2929-0.0978\imo$&$0.299-0.101\imo$\\
$0.02$&$0.2603-0.0911\imo$&$0.263-0.093\imo$\\
$0.04$&$0.2247-0.0821\imo$&$0.226-0.083\imo$\\
$0.06$&$0.1854-0.0701\imo$&$0.187-0.071\imo$\\
$0.08$&$0.1404-0.0542\imo$&$0.141-0.055\imo$\\
$0.09$&$0.11392-0.04397\imo$&$0.1147-0.0443\imo$\\
$0.10$&$0.08156-0.03121\imo$&$0.0819-0.0315\imo$\\
$0.11$&$0.02549-0.00965\imo$&$0.02550-0.00967\imo$\\
\hline
\end{longtable}
\end{minipage}
\begin{minipage}[c]{9cm}
\centerline{\MSF: $l=2$, $n=0$}
\begin{longtable}{|r|r|r|}
\hline
$\L$ & $\om$ (WKB) & $\om$ (P-T)\\
\hline
   $0$&$0.48364-0.09677\imo$&$0.487-0.098\imo$\\
$0.02$&$0.43461-0.08858\imo$&$0.437-0.089\imo$\\
$0.04$&$0.38078-0.07876\imo$&$0.382-0.079\imo$\\
$0.06$&$0.32002-0.06685\imo$&$0.321-0.067\imo$\\
$0.08$&$0.24747-0.05197\imo$&$0.248-0.052\imo$\\
$0.09$&$0.20296-0.04256\imo$&$0.2033-0.0427\imo$\\
$0.10$&$0.14661-0.03069\imo$&$0.1468-0.0308\imo$\\
$0.11$&$0.04617-0.00963\imo$&$0.04617-0.00963\imo$\\
\hline
\end{longtable}
\end{minipage}
\bigskip\newline
\begin{minipage}[c]{9cm}
\centerline{\MSF: $l=2$, $n=1$}
\begin{longtable}{|r|r|r|}
\hline
$\L$ & $\om$ (WKB) & $\om$ (P-T)\\
\hline
   $0$&$0.46385-0.29563\imo$&$0.487-0.294\imo$\\
$0.02$&$0.42084-0.26862\imo$&$0.437-0.268\imo$\\
$0.04$&$0.37165-0.23796\imo$&$0.382-0.238\imo$\\
$0.06$&$0.31422-0.20163\imo$&$0.321-0.202\imo$\\
$0.08$&$0.24426-0.15627\imo$&$0.248-0.156\imo$\\
$0.09$&$0.20098-0.12794\imo$&$0.2033-0.1282\imo$\\
$0.10$&$0.14576-0.09212\imo$&$0.1468-0.0923\imo$\\
$0.11$&$0.04614-0.02889\imo$&$0.04617-0.02890\imo$\\
\hline
\end{longtable}
\end{minipage}
\def\EMF{\lang{QNMs of electromagnetic field}{Частоти електромагнітного поля}}
\begin{minipage}[c]{9cm}
\centerline{\EMF: $l=1$, $n=0$}
\begin{longtable}{|r|r|r|}
\hline
$\L$ & $\om$ (WKB) & $\om$ (P-T)\\
\hline
   $0$&$0.2482-0.0926\imo$&$0.255-0.096\imo$\\
$0.02$&$0.2259-0.0842\imo$&$0.231-0.087\imo$\\
$0.04$&$0.2006-0.0748\imo$&$0.204-0.077\imo$\\
$0.06$&$0.1709-0.0639\imo$&$0.173-0.065\imo$\\
$0.08$&$0.1339-0.0502\imo$&$0.135-0.051\imo$\\
$0.09$&$0.11053-0.04156\imo$&$0.1110-0.0419\imo$\\
$0.10$&$0.08035-0.03028\imo$&$0.0805-0.0304\imo$\\
$0.11$&$0.02545-0.00962\imo$&$0.02546-0.00962\imo$\\
\hline
\end{longtable}
\end{minipage}
\bigskip\newline
\begin{minipage}[c]{9cm}
\centerline{\EMF: $l=2$, $n=0$}
\begin{longtable}{|r|r|r|}
\hline
$\L$ & $\om$ (WKB) & $\om$ (P-T)\\
\hline
   $0$&$0.45759-0.09501\imo$&$0.461-0.096\imo$\\
$0.02$&$0.41502-0.08615\imo$&$0.418-0.087\imo$\\
$0.04$&$0.36723-0.07624\imo$&$0.369-0.077\imo$\\
$0.06$&$0.31182-0.06478\imo$&$0.313-0.065\imo$\\
$0.08$&$0.24365-0.05067\imo$&$0.244-0.051\imo$\\
$0.09$&$0.20085-0.04180\imo$&$0.2012-0.0419\imo$\\
$0.10$&$0.14582-0.03037\imo$&$0.1459-0.0304\imo$\\
$0.11$&$0.04614-0.00962\imo$&$0.04615-0.00962\imo$\\
\hline
\end{longtable}
\end{minipage}
\begin{minipage}[c]{9cm}
\centerline{\EMF: $l=2$, $n=1$}
\begin{longtable}{|r|r|r|}
\hline
$\L$ & $\om$ (WKB) & $\om$ (P-T)\\
\hline
   $0$&$0.43653-0.29073\imo$&$0.461-0.289\imo$\\
$0.02$&$0.39900-0.26202\imo$&$0.418-0.261\imo$\\
$0.04$&$0.35602-0.23065\imo$&$0.369-0.231\imo$\\
$0.06$&$0.30498-0.19516\imo$&$0.313-0.196\imo$\\
$0.08$&$0.24046-0.15223\imo$&$0.244-0.153\imo$\\
$0.09$&$0.19907-0.12549\imo$&$0.2012-0.1258\imo$\\
$0.10$&$0.14515-0.09114\imo$&$0.1459-0.0913\imo$\\
$0.11$&$0.04612-0.02886\imo$&$0.04615-0.02887\imo$\\
\hline
\end{longtable}
\end{minipage}
\def\GF{\lang{QNMs of gravitational field}{Частоти гравітаційного поля}}
\bigskip\newline
\begin{minipage}[c]{18cm}
\centerline{\GF: $l=2$, $n=0$}
\begin{longtable}{|l|r|r|r|r|}
\hline
$\L$ & $\om$ (WKB) (odd) & $\om$ (P-T) (odd) & $\om$ (WKB) (even) & $\om$ (P-T) (even)\\
\hline
   $0$&$0.3736-0.0889\imo$&$0.378-0.091\imo$&$0.3737-0.0889\imo$&$0.378-0.091\imo$\\
$0.02$&$0.3384-0.0817\imo$&$0.342-0.083\imo$&$0.3384-0.0817\imo$&$0.342-0.083\imo$\\
$0.04$&$0.2989-0.0733\imo$&$0.301-0.074\imo$&$0.2989-0.0733\imo$&$0.301-0.074\imo$\\
$0.06$&$0.2533-0.0630\imo$&$0.254-0.064\imo$&$0.2533-0.0630\imo$&$0.254-0.064\imo$\\
$0.08$&$0.1975-0.0499\imo$&$0.198-0.050\imo$&$0.1975-0.0499\imo$&$0.198-0.050\imo$\\
$0.09$&$0.16261-0.04136\imo$&$0.1629-0.0415\imo$&$0.16261-0.04137\imo$&$0.1629-0.0415\imo$\\
$0.10$&$0.11792-0.03021\imo$&$0.1180-0.0303\imo$&$0.11792-0.03021\imo$&$0.1180-0.0303\imo$\\
$0.11$&$0.03727-0.00962\imo$&$0.03727-0.00962\imo$&$0.03727-0.00962\imo$&$0.03727-0.00962\imo$\\
\hline
\end{longtable}
\end{minipage}
\bigskip\newline
\begin{minipage}[c]{18cm}
\centerline{\GF: $l=2$, $n=1$}
\begin{longtable}{|r|r|r|r|r|}
\hline
$\L$ & $\om$ (WKB) (odd) & $\om$ (P-T) (odd) & $\om$ (WKB) (even) & $\om$ (P-T) (even)\\
\hline
   $0$&$0.3463-0.2735\imo$&$0.378-0.272\imo$&$0.3467-0.2739\imo$&$0.378-0.272\imo$\\
$0.02$&$0.3185-0.2488\imo$&$0.342-0.249\imo$&$0.3188-0.2491\imo$&$0.342-0.249\imo$\\
$0.04$&$0.2858-0.2215\imo$&$0.301-0.223\imo$&$0.2858-0.2217\imo$&$0.301-0.223\imo$\\
$0.06$&$0.2457-0.1896\imo$&$0.254-0.191\imo$&$0.2457-0.1898\imo$&$0.254-0.191\imo$\\
$0.08$&$0.1941-0.1497\imo$&$0.198-0.150\imo$&$0.1941-0.1498\imo$&$0.198-0.150\imo$\\
$0.09$&$0.16079-0.12412\imo$&$0.1629-0.1246\imo$&$0.16079-0.12415\imo$&$0.1629-0.1246\imo$\\
$0.10$&$0.11724-0.09063\imo$&$0.1180-0.0908\imo$&$0.11724-0.09064\imo$&$0.1180-0.0908\imo$\\
$0.11$&$0.03725-0.02885\imo$&$0.03727-0.02885\imo$&$0.03725-0.02885\imo$&$0.03727-0.02885\imo$\\
\hline
\end{longtable}
\end{minipage}
\bigskip\newline
\begin{minipage}[c]{18cm}
\centerline{\GF: $l=3$, $n=0$}
\begin{longtable}{|r|r|r|r|r|}
\hline
$\L$ & $\om$ (WKB) (odd) & $\om$ (P-T) (odd) & $\om$ (WKB) (even) & $\om$ (P-T) (even)\\
\hline
   $0$&$0.599443-0.092703\imo$&$0.602-0.093\imo$&$0.599443-0.092703\imo$&$0.602-0.093\imo$\\
$0.02$&$0.543115-0.084495\imo$&$0.545-0.085\imo$&$0.543115-0.084496\imo$&$0.545-0.085\imo$\\
$0.04$&$0.480058-0.075146\imo$&$0.481-0.076\imo$&$0.480058-0.075146\imo$&$0.481-0.076\imo$\\
$0.06$&$0.407175-0.064140\imo$&$0.408-0.064\imo$&$0.407175-0.064140\imo$&$0.408-0.064\imo$\\
$0.08$&$0.317805-0.050382\imo$&$0.318-0.051\imo$&$0.317804-0.050382\imo$&$0.318-0.051\imo$\\
$0.09$&$0.261841-0.041644\imo$&$0.2621-0.0417\imo$&$0.261843-0.041643\imo$&$0.2621-0.0417\imo$\\
$0.10$&$0.189994-0.030314\imo$&$0.1901-0.0303\imo$&$0.189994-0.030314\imo$&$0.1901-0.0303\imo$\\
$0.11$&$0.060091-0.009619\imo$&$0.06009-0.00962\imo$&$0.060091-0.009619\imo$&$0.06009-0.00962\imo$\\
\hline
\end{longtable}
\end{minipage}
\bigskip\newline
\begin{minipage}[c]{18cm}
\centerline{\GF: $l=3$, $n=1$}
\begin{longtable}{|r|r|r|r|r|}
\hline
$\L$ & $\om$ (WKB) (odd) & $\om$ (P-T) (odd) & $\om$ (WKB) (even) & $\om$ (P-T) (even)\\
\hline
   $0$&$0.58264-0.28129\imo$&$0.602-0.280\imo$&$0.58264-0.28129\imo$&$0.602-0.280\imo$\\
$0.02$&$0.53074-0.25536\imo$&$0.545-0.255\imo$&$0.53074-0.25536\imo$&$0.545-0.255\imo$\\
$0.04$&$0.47166-0.22639\imo$&$0.481-0.227\imo$&$0.47166-0.22639\imo$&$0.481-0.227\imo$\\
$0.06$&$0.40217-0.19281\imo$&$0.408-0.193\imo$&$0.40217-0.19281\imo$&$0.408-0.193\imo$\\
$0.08$&$0.31550-0.15125\imo$&$0.318-0.152\imo$&$0.31549-0.15125\imo$&$0.318-0.152\imo$\\
$0.09$&$0.26056-0.12498\imo$&$0.2621-0.1251\imo$&$0.26057-0.12497\imo$&$0.2621-0.1251\imo$\\
$0.10$&$0.18952-0.09095\imo$&$0.1901-0.0910\imo$&$0.18952-0.09095\imo$&$0.1901-0.0910\imo$\\
$0.11$&$0.06008-0.02886\imo$&$0.06009-0.02886\imo$&$0.06008-0.02886\imo$&$0.06009-0.02886\imo$\\
\hline
\end{longtable}
\end{minipage}
\def\MDF{\lang{QNMs of massless Dirac field}{Частоти безмасового поля Дірака}}
\bigskip\newline
\begin{minipage}[c]{18cm}
\centerline{\MDF: $\k=1$, $n=0$}
\begin{longtable}{|r|r|r|r|r|}
\hline
$\L$ & $\om$ (WKB) (-)& $\om$ (P-T) (-)& $\om$ (WKB) (+) & $\om$ (P-T) (+)\\
\hline
   $0$&$0.183-0.097\imo$&$0.183-0.098\imo$&$0.183-0.095\imo$&$0.189-0.105\imo$\\
$0.02$&$0.167-0.087\imo$&$0.166-0.088\imo$&$0.167-0.085\imo$&$0.171-0.094\imo$\\
$0.04$&$0.149-0.077\imo$&$0.147-0.077\imo$&$0.149-0.075\imo$&$0.151-0.082\imo$\\
$0.06$&$0.128-0.069\imo$&$0.127-0.065\imo$&$0.128-0.064\imo$&$0.128-0.069\imo$\\
$0.08$&$0.1005-0.0507\imo$&$0.0985-0.0505\imo$&$0.1006-0.0501\imo$&$0.1001-0.0529\imo$\\
$0.09$&$0.0832-0.0418\imo$&$0.0813-0.0415\imo$&$0.0833-0.0414\imo$&$0.0824-0.0432\imo$\\
$0.10$&$0.0606-0.0304\imo$&$0.0592-0.0301\imo$&$0.0606-0.0303\imo$&$0.0597-0.0311\imo$\\
$0.11$&$0.0192-0.0096\imo$&$0.0188-0.0096\imo$&$0.0192-0.0096\imo$&$0.0189-0.0097\imo$\\
\hline
\end{longtable}
\end{minipage}
\bigskip\newline
\begin{minipage}[c]{18cm}
\centerline{\MDF: $\k=2$, $n=0$}
\begin{longtable}{|r|r|r|r|r|}
\hline
$\L$ & $\om$ (WKB) (-)& $\om$ (P-T) (-)& $\om$ (WKB) (+) & $\om$ (P-T) (+)\\
\hline
   $0$&$0.38004-0.09642\imo$&$0.382-0.097\imo$&$0.38007-0.09637\imo$&$0.386-0.099\imo$\\
$0.02$&$0.34491-0.08715\imo$&$0.346-0.088\imo$&$0.34493-0.08712\imo$&$0.349-0.089\imo$\\
$0.04$&$0.30542-0.07690\imo$&$0.306-0.077\imo$&$0.30543-0.07688\imo$&$0.308-0.079\imo$\\
$0.06$&$0.25953-0.06516\imo$&$0.260-0.065\imo$&$0.25954-0.06515\imo$&$0.261-0.066\imo$\\
$0.08$&$0.20296-0.05085\imo$&$0.203-0.051\imo$&$0.20296-0.05084\imo$&$0.204-0.052\imo$\\
$0.09$&$0.16738-0.04190\imo$&$0.167-0.0418\imo$&$0.16738-0.04190\imo$&$0.168-0.0423\imo$\\
$0.10$&$0.12157-0.03041\imo$&$0.121-0.0304\imo$&$0.12157-0.03041\imo$&$0.122-0.0305\imo$\\
$0.11$&$0.03849-0.00962\imo$&$0.0384-0.0096\imo$&$0.03849-0.00962\imo$&$0.0384-0.0096\imo$\\
\hline
\end{longtable}
\end{minipage}
\bigskip

\lang{The relative error of the WKB method depends significantly on a field under consideration. Comparison of the sixth and fifth WKB order let us to conclude that the worst WKB convergence gives the Dirac field, where the expected error reaches 1\% in the sixth WKB order. The results calculated in \cite{Cho} by using the third WKB order differ from the sixth order values up to 4\%. Thus, the possible relative error of the third order WKB results may exceed 4\%.}
{Відносна похибка методу ВКБ істотно залежить від поля, що розглядається. Порівняння шостого і п'ятого порядків дозволяє зробити висновок, що найгірше формула ВКБ збігається для Діраківського поля, де похибка оцінюється в 1\% для шостого порядку. Результати обчислення в \cite{Cho} за допомогою третього порядку ВКБ відрізняються від результатів шостого порядку до 4\%, тому можлива відносна похибка третього порядку ВКБ може перевищувати 4\%.}

\lang{We can also evaluate the relative error order for gravitational modes by comparison QNMs of odd and even perturbations. Because the potentials can be expressed in the form}
{Відносну похибку для гравітаційних мод, можна оцінити також порівнюючи частоти, що відповідають аксіальним та полярним збуренням. Оскільки відповідні потенціали можуть бути представлені у вигляді}
\be\label{goec}
V_{odd}= W_g^2 + \frac{dW_g}{dr^*}+\b; \qquad V_{even} = W_g^2 -
\frac{dW_g}{dr^*}+\b;
\ee
\lang{where}{де}
$$W_g = \frac{2M}{r^2}-\frac{3+2c}{3r}+\frac{3c^2+2c^3-9\L M^2}{3c(3M+cr)}-\frac{1}{3M}(c^2+c-\frac{3\L M^2}{c}); \qquad \b = -\frac{c^2(c+1)^2}{9M^2};$$
\lang{any solution for even parity can be found from the solution for odd parity}
{кожний розв'язок для парних збурень можна знайти з відповідного розв'язку для непарних збурень}
\cite{Chandrasekhar}:
$$\P_{even}(r^*)=q\left(W_g(r^*) - \frac{d}{dr^*}\right)\P_{odd}(r^*); \qquad q = const.$$
\lang{Obviously, if a solution $\P_{odd}$ describing quasi-normal oscillations satisfies the boundary conditions \p{bounds}, the $\P_{even}$ solution satisfies the same ones describing
quasi-normal oscillations too.}
{Очевидно, що якщо розв'язок $\P_{odd}$, що описує квазинормальні коливання задовольняє граничні умови \p{bounds}, розв'язок $\P_{even}$ також задовольняє \p{bounds}.}

\lang{The same fact \cite{Potentials} takes place for massless Dirac QNMs of opposite chiralities}
{Такий самий факт \cite{Potentials} має місце для квазинормальних мод протилежної спіральності безмасового Діраківського поля}:
\be\label{docc}
V_d = W_d^2 \pm \frac{dW_d}{dr^*}; \qquad W_d(r) =
\frac{\k\sqrt{f(r)}}{r}.
\ee

\lang{Therefore, the QNM spectrum is the same for odd and even gravitational perturbations and for opposite chirality of Dirac perturbations in SdS background as well as in asymptotically flat universe. We see that the difference between the numerical results does not exceed our expectations.}
{Через це, на тлі Шварцшильда-де Сітера, так само, як і на асимптотично плоскому тлі, спектр квазинормальних частот є однаковим для аксіальних і полярних збурень гравітаційного поля, а також збурень Діраківського поля протилежної спіральності. Легко бачити, що різниця між чисельними значеннями не перевищує очікуваної похибки.}

\lang{We can see that the imaginary parts of QNMs, calculated using both sixth-order WKB and Pöshl-Teller approaches, depend on multipole and overtone number and black hole mass only near the extremal value of $\L$. This agrees with \cite{Cardoso-Lemos2}, where the QNMs of near extremal SdS space-time were found}
{Можна бачити, що уявна частина квазинормальних частот, обчислених за допомогою шостого порядку ВКБ і підходу Пьошля-Теллера, біля екстремального значення $\L$ залежить тільки від номера овертону і мультиполя, а також від маси чорної діри. Це узгоджується з \cite{Cardoso-Lemos2}, де знайдено квазинормальні частоти для майже екстремальних чорних дір}:
\be\label{ese}
\om b = -\left(n+\frac 1 2\right)\imo + \sqrt{l(l+1)-\frac 1
4}
\ee
\lang{for scalar and electromagnetic perturbations,}{для збурень скалярного та електромагнітного поля і}
\be\label{eg}
\om b = -\left(n+\frac 1 2\right)\imo + \sqrt{(l+2)(l-1)-\frac 1
4}
\ee
\lang{for gravitational perturbations. Here}{для гравітаційного. Тут}
$$b = \frac{54M^3}{(r_c-3M)(r_c + 6M)}, \qquad r_c\to 3M.$$
\lang{Using the same technique, one can find that}{Користуючись такою самою технікою, можна знайти, що}
\be\label{eds}
\om b = -\left(n+\frac 1 2\right)\imo + \sqrt{\k^2-\frac 1
4}
\ee
\lang{for massless Dirac field perturbations. It means in the extremal limit fields of different spin decay with the same rate. Note, that formulae \p{ese}, \p{eg}, \p{eds} give right results for any overtone number $n$ and may be useful for studying the $n\rightarrow\infty$ limit.}
{для збурень безмасового Діраківського поля. Це означає, що в екстремальній границі поля різного спіну спадають з однаковою швидкістю. Зауважимо, що формули \p{ese}, \p{eg}, \p{eds} є правильними для будь-якого номера овертона $n$ і можуть виявитися корисними для вивчення границі $n\rightarrow\infty$.}

\lang{The well-known approximate formula for large multipole number $l$ (or $\k$ in \p{mdp})}
{Відома наближена формула для великих значеннь номера мультиполя $l$ (або $\k$ у \p{mdp})}
\cite{FerrariMashhoon,Press}:
\bea\label{aflls}
\om &=& \frac{1}{3\sqrt 3 M}\left(l + \frac 1 2 - \left(n + \frac 1 2\right)\imo\right) + O\left(\frac 1 l\right);
\\\nonumber\om &=& \frac{1}{3\sqrt 3 M}\left(\k - \left(n + \frac 1 2\right)\imo\right) + O\left(\frac 1 \k\right);
\eea
\lang{can be generalized for the case of non-zero $\L$. Making use of the first-order WKB method or the formula of the approximation by the Pöshl-Teller potential one can find for the non-extremal SdS black hole}
{може бути узагальнена на випадок ненульових $\L$. Користуючись першим порядком формули ВКБ або наближенням потенціалом Пьошля-Теллера, можна знайти, що для неекстремальної чорної діри Шварцшильда-де Сітера}:
\bea\label{afllads}
\om &=& \frac{\sqrt{1 - 9M^2\L}}{3\sqrt 3 M}\left(l + \frac 1 2 - \left(n + \frac 1 2\right)\imo\right) + O\left(\frac 1 l\right);
\\\nonumber\om &=& \frac{\sqrt{1 - 9M^2\L}}{3\sqrt 3 M}\left(\k - \left(n + \frac 1 2\right)\imo\right) + O\left(\frac 1 \k\right).
\eea

\section{\lang{Conclusion}{Заключення}}
\lang{The QNMs of Schwarzschild-de Sitter black holes for fields of different spin have been calculated. The modes are determined by the black hole mass $M$ and the cosmological constant $\L$ only. The frequencies all have a negative imaginary part, which means that the black hole is stable against these perturbations. The presence of the cosmological constant leads to decrease of the real oscillation frequency and to a slower decay. The interesting problem that was outside our consideration is the search of asymptotic behavior of SdS QNMs at large imaginary part ($n \rightarrow \infty$) for which other approaches should be explored
\cite{Nollert}.}
{Були обчислені квазинормальні моди чорної діри Шварцшильда-де Сітера. Ці моди визначаються тільки масою чорної діри $M$ і космологічною сталою $\L$. Усі частоти мають від'ємну уявну частину, що означає стабільність чорної діри відносно розглянутих збурень. Наявність космологічної сталої призводить до зменшення дійсної частоти коливань і більш повільного згасання. Цікавою проблемою, яка залишилася поза розглядом, є пошук асимптотичної поведінки квазинормальних мод у просторі Шварцшильда-де Сітера для великої уявної частини ($n \rightarrow \infty$), для розв'язку якої слід використовувати інші підходи \cite{Nollert}.}

\section*{\lang{Acknowledgements}{Подяка}}
\lang{I would like to thank R. Konoplya for proposing this problem to me and reading the manuscript.}
{Хочу подякувати Роману Коноплі за запропоновану задачу і перевірку цієї статті.}

\end{document}